\documentclass[preprint2]{aastex}
\usepackage{amssymb}
\usepackage{graphicx}

\shorttitle{ Statistically optimal fitting of astrometric signals }
\shortauthors{ Gai, Busonero and Cancelliere }

\begin{document}

\title{ Statistically optimal fitting of astrometric signals }

\author{M. Gai\altaffilmark{1} and D. Busonero\altaffilmark{1}}
\affil{ INAF - Osservatorio Astrofisico di 
Torino, V. Osservatorio 20, 10025 Pino T.se (TO), Italy }
\email{gai@oato.inaf.it}

\and

\author{R. Cancelliere\altaffilmark{2}}
\affil{Universit\`a di Torino, Dipartimento di Informatica, 
C.so Svizzera 185, 10149 Torino, Italy}

\begin{abstract}
A general purpose fitting model for one-dimensional astrometric 
signals is developed, building on a maximum likelihood framework, 
and its performance is evaluated by simulation over a set 
of realistic image instances. 
The fit quality is analysed as a function of the number 
of terms used for signal expansion, and of astrometric error, 
rather than RMS discrepancy with respect to the input signal. 
The tuning of the function basis to the statistical characteristics 
of the signal ensemble is discussed. The fit sensitivity to a priori 
knowledge of the source spectra is addressed. 
Some implications of the current results on calibration and data 
reduction aspects are discussed, in particular with respect to 
Gaia. 
\end{abstract}


\section{Introduction}
\label{sec:Introduction}
For efficient extraction of astrometric and photometric information 
from astronomical data, we often rely on an effective mathematical 
representation of the signal, suitable to the 
processing and analysis activities \citep{Mighell}. 
The issue was evidenced e.g. by the attempts at image 
deconvolution of the initial HST images 
\citep{KristHasan1993}. 
The need for modeling the Point Spread Function (PSF) 
variation, in particular over the large field of view of some 
modern instruments, leads to different techniques according 
to specific application needs 
\citep{LSV_PSF2008}. 
The dependence of astrometry from source color 
(i.e. spectral distribution) must be taken into account, even 
for all reflective optics in space 
\citep{1984SPIE..445..497L,2010SPIE.7734E.144N}. 
Also on ground usage of adaptive optics, which recovers up to 
a point the diffraction limited performance, requires a detailed 
model of the PSF for precision measurements 
\citep{Britton2006}. 
Modeling and calibration of images for differential astrometry 
has been addressed from both mathematical and experimental 
standpoints, demonstrating micro-pixel precision 
\citep{Chengxing}. 

In modern spaceborne instruments, with complex telescopes 
and large area imaging detectors, the instrument response 
changes over the focal plane, or over the corresponding 
field of view. 
The change is due to both optics and detector, related 
to the geometry and (electro-)optical parameter variation. 
The effect is both local, modifying the detected signal 
with respect to the case of diffraction limited images 
onto an ideal detector, and global, affecting the 
projection of the focal plane geometry onto the sky. 
The monochromatic PSF variation is usually attributed to 
the optical system, described in terms of wavefront error (WFE), 
and for measurements over a finite bandwidth the signal is weighted 
by the source spectral distribution. 
Realistic detector characteristics affect the signal 
distribution, e.g. because of its geometry and the 
Modulation Transfer Function (MTF); it is therefore possible 
to describe the detected signal shape in terms of an 
equivalent wavefront, inducing a comparable overall effect. 
However, detector parameters may have sharp variation over the field, 
even from pixel to pixel. 
Besides, field distortion is a large scale effect, introducing 
smooth variation of the ratio between given small angles on the 
sky and their projected linear value on the focal plane 
(e.g. with typical ``barrel" or ``pincushion" distribution). 

Both local and large scale effects can be made acceptably small 
for many applications by proper design and implementation, 
but in order to preserve the ultimate measurement performance 
of an instrument we have to account for them by calibration either 
on the sky, or through adequate subsystems (e.g. metrology). 

This paper deals with the issue of a convenient 
mathematical representation of the profile variation, 
over the focal plane and throughout operation, of 
realistic polychromatic images from a range of objects, 
in the perspective of a modular approach to data reduction 
and calibration. 
For simplicity, we deal with one-dimensional signals; 
the method can be applied to conventional images e.g. 
separately along each axis, in principle. 

We previously addressed the signal modeling issue, focusing on 
application to Gaia, in \citet{2010MNRAS.406.2433G}. 
Here we describe further results of our investigation aimed at 
strengthening the conceptual framework and improving on specific 
implementation related aspects. 
The goal is to define a compact, computationally efficient 
representation of the signal profile, over finite focal plane 
regions, suitable to the processing of data from stars 
spanning a range of magnitude and spectral type. 
Therefore, the fit optimization is statistical, i.e. focused on 
collective rather than individual performance. 

It is assumed that calibration of field distortion and detector 
parameter is performed separately from that of the local signal 
``shape", although in practice the implementation of one module 
may affect the performance of others. 
Such issues, therefore, are not discussed in the current work. 
It may be remarked that at least some detector parameter variations, 
e.g. smooth changes of MTF and charge transfer efficiency (CTE), 
may be easily encoded by their effects on the signal profile using 
the proposed modeling tool. 

In our approach, the spatially variant PSF is expanded in a sum 
of spatially invariant functions, with coefficients varying over 
the field to describe the instrument response variation for given 
spectral distribution sources, as in \citet{Lauer2002}. 
The fitting process follows the maximum likelihood approach, 
with some allowance: 
on one side, the fit must be compliant with a reasonable range 
of astrophysical parameter variation among sources; 
besides, it should be tolerant with respect to instrument 
response evolution along operations. 

In Sec.~\ref{sec:Astrometric-fit-error} we deal with
the mathematical formulation of an astrometrically sound 
fitting process, minimizing the photo-center discrepancy 
between input signal and output fit. 
This should minimize the systematic errors introduced 
in applying the fit result as a template, calibrated on 
a given data set, for location estimate on new data. 

Then, in Sec.~\ref{sec:Function-basis}, the fitting functions 
are defined as an orthogonal set, in order to  minimize  
noise amplification effects and / or correlations in the data 
reduction results \citep{Makarov2012}. 

We evaluate the feasibility and performance of our method 
by simulation on a study case consistent with the Gaia 
framework, described in Sec.~\ref{sec:GaiaSig}, although 
the tuning of the fitting method to other applications appears 
to be straightforward. 
The approach to calibration 
of detector and field distortion in Gaia will also be briefly 
discussed.

The data sets and the fitting algorithm performance are discussed
in Sec.~\ref{sec:Simulations1}, evaluating the fit discrepancy with
respect to the input signals from the standpoint of photometric and
astrometric error as a function of the model parameters. 
The impact on implementation concerns the number of fitting terms 
required, within a given range of astrophysical and instrumental 
parameter variation, and of model related location error. 

In Sec.~\ref{sec:Spectral-sensitivity} we evaluate the sensitivity
of the proposed fit model to the natural variation of spectral 
distribution among stars. 
In particular, the location error associated to a given spectral 
mismatch between the signal template (i.e. the fit result) and 
new data sets is evaluated. 
The investigation on fit tolerance to the knowledge of the source 
effective temperature is relevant to practical implementation within 
the data reduction system, since it defines the number of different 
templates required over the whole spectral range to preserve an 
adequate precision level. 

Finally, in Sec.~\ref{sec:Discussion}, we discuss the potential impact 
of our results on calibration and implementation aspects, and then 
we draw our conclusions.

\section{Astrometric error of the fit}
\label{sec:Astrometric-fit-error}
The detected signal, sampled on $K$ pixels centered on positions $x_{k},\, k=1,\ldots,K$,
is a function $f\left(x_{k}-\tau_{T}\right)$, located at the ``true''
photo-center position $\tau_{T}$, which is the input to the
fitting process. 

It should be noted that our ``signal'' is {\em not} coincident with 
the readout from the elementary exposure of a single star, even a 
bright one, but rather the synthetic representation of the instrument 
response, for a given set of conditions, at very high resolution, 
as in \citet{Guillard2010}. 
It may be derived from the measured data e.g. by combination of a 
large set of elementary exposures on a given CCD, in order to achieve 
a very high signal to noise ratio (SNR) and oversample the instrument 
response, taking advantage of the natural variation in the relative 
phase between individual stars and the detector. 
In practice, the signal used for the fit must have an equivalent 
noise much below that of any individual exposure (at the $\mu as$ 
level or below) to  minimize the effects of model errors in its 
subsequent usage. 
Residual elementary exposure errors, e.g. from individual 
background subtraction, are supposed to be reduced by averaging 
over the whole data set. 
Also, implementation will require suitable composition rules e.g. 
for magnitude dependent weighting and consistency check. 

Let us assume that $\tilde{f}\left(x_{k}\right)$, i.e. the output
of the fitting process, is an approximation of the input function
$f$, with a residual labeled hereafter as the \textit{fit discrepancy}
$h\left(x_{k}\right)$: 
\begin{equation}
f\left(x_{k}-\tau_{T}\right) = 
\tilde{f}\left(x_{k}\right)+h\left(x_{k}\right)\,.
\label{eq:FitError}
\end{equation}
It may be noted that a general purpose fit is usually aimed at 
minimizing the discrepancy $h\left(x_{k}\right)$, e.g. in the 
least square sense, whereas our goal is to  minimize the astrometric 
error $\delta\tau$ experienced when using the fit result 
$\tilde{f}\left(x_{k}\right)$ as a template within the location 
algorithm on new sets of measurements distributed according 
to $f$. 
Although obviously the two requirements coincide in the limit of 
small errors ($\delta\tau\rightarrow0\,\Leftrightarrow\, 
h\left(x_{k}\right)\rightarrow0$), 
the fit with a limited number of terms may have different 
performance according to either criteria.

In this paper, we address the issue of signal fitting with 
the focus on minimization of the systematic \textit{astrometric 
error} or bias. 

The Maximum Likelihood photo-center estimate builds on 
the approach described in \citet{1998PASP..110..848G},
recalled below. 
The fit discrepancy $h$ can be represented in that framework as 
a measurement error, 
introducing a photo-center location error when the optimal location 
estimate $\tau_{E}$ is searched for by minimization of a weighted 
square error functional derived from the classical definition 
of $\chi^{2}$: 
\begin{equation}
\chi^{2} = 
\sum_{k}\frac{\left[f_{k}-\tilde{f}_{k}\right]^{2}}{\sigma_{k}^{2}} 
\,.\label{eq:ChiSquare}
\end{equation}
The variable dependence here is implicit to simplify the notation: $f\left(x_{k}\right)=f_{k}$,
and so on. 
The photo-center location estimate is thus independent from the 
Center Of Gravity (COG) algorithm, which is known to be less than 
optimal and potentially affected by systematic error, i.e. an 
astrometric \textit{bias} \citep{1978moas.coll..197L}. 

The variance $\sigma^{2}\left(x_{k}\right)$ associated to the 
signal distribution $f$ is assumed to be known. 
It will in general account for several noise sources (signal and 
background photon statistics, readout noise, dark current, and 
so on), 
with the best case limited at the very least by shot noise: 
$\sigma^{2}\left(x_{k}\right)\geq f\left(x_{k}\right)\geq0$ 
(with all quantities scaled to photons for simplicity). 

The location process must find the stationary points of 
Eq.~\ref{eq:ChiSquare}, i.e. solve the equation 
$\frac{d\chi^{2}}{d\tau}=0$, 
which may be expanded at first order to 
\begin{equation}
\sum_{k}\frac{\left[f_{k}-\tilde{f}_{k}\right]f_{k}'}{\sigma_{k}^{2}} 
= 0\,,\label{eq:ZeroSearch}
\end{equation}
assuming small errors, in particular the astrometric 
error $\delta\tau=\tau_{E}-\tau_{T}$, so that e.g. 
$f\left(x_{k}-\tau_{E}\right)\simeq f\left(x_{k}-\tau_{T}\right)$. 

Therefore, taking advantage of the Taylor's expansion of the signal
model, i.e. that 
\begin{equation}
f\left(x_{k}-\tau_{E}\right)\simeq f\left(x_{k}-\tau_{T}\right)-\delta\tau\cdot f\,'\left(x_{k}-\tau_{T}\right)\,,\label{eq:Taylor1}
\end{equation}
we get the location estimate error as 
\begin{equation}
\tau_{E}-\tau_{T}=-\frac{\sum_{k}\, h\left(x_{k}\right)f\,'\left(x_{k}\right)/\sigma_{k}^{2}}{\sum_{k}\,\left[f\,'\left(x_{k}\right)\right]^{2}/\sigma_{k}^{2}}\,.\label{eq:BiasEstim}
\end{equation}

The location error introduced by the fit may be unbiased, i.e. with 
zero expected value, under the assumption that the fit discrepancy 
is itself characterized by zero mean. 
In general, the correlation between pixels may not be zero, so that 
the variance estimation of the location error is not trivial. 
However, it may be remarked that, through Eq.~\ref{eq:BiasEstim}, 
it is possible not only to assess the fit quality by providing an 
estimate of the astrometric bias introduced by the fitting process, 
correct from the standpoint of the maximum likelihood approach, but 
in principle to correct such error as well. 

It may be noted that the $\chi^{2}$ formulation, through 
Eq.~\ref{eq:ChiSquare}, is strictly related to the individual 
signal distribution.

\section{Function basis}
\label{sec:Function-basis}
The parent function proposed for generation of the monochromatic 
basis functions 
is the squared $sinc$ function, depending on an adimensional argument 
related to 
the focal plane coordinate $x$, the wavelength $\lambda$ and the 
characteristic width $L_{\xi}$, as 
\begin{equation}
\psi_{0}^{m}\left(x\right) = sinc^{2}\rho = 
\left[\frac{\sin\rho}{\rho}\right]^{2},\; 
\rho = \pi\frac{xL_{\xi}}{\lambda F}\,. 
\label{eq:Parent_function0}
\end{equation}
This corresponds to the signal generated by a rectangular infinite
slit having width $L_{\xi}$, in the ideal (aberration free) case of a 
telescope with effective focal length $F$. 
The contribution of finite pixel size, nominal modulation transfer 
function (MTF) and CCD operation in Time-Delay Integration (TDI) 
mode are also included. 
Higher order functions are generated by suitable combinations
of the parent function and its derivatives 
\begin{equation}
\psi_{n}^{m}\left(x\right) = 
\frac{d}{dx}\psi_{n-1}^{m}\left(x\right) = 
\left(\frac{d}{dx}\right)^{n}\psi_{0}^{m}\left(x\right) \, , 
\label{eq:Base_Functions}
\end{equation}
according to a construction rule ensuring ortho-normality by integration 
over the domain, i.e. 
\begin{equation}
\left\langle \psi_{p}^{m}\,\psi_{q}^{m}\right\rangle = 
\sum_{k}\,\psi_{p}^{m}\left(x_{k}\right)\,\psi_{q}^{m}\left(x_{k}\right) = 
\delta_{pq},\, 
\label{eq:OrthoNorm}
\end{equation}
using the Kronecker's $\delta$ notation. 

The polychromatic functions are built according to linear superposition
of the monochromatic functions, weighted by the normalized detected
source spectrum $S$ (which includes the system response), here 
discretized and indexed by $l$: 
\begin{equation}
\psi_{n}\left(x\right) = 
\sum_{l}S_{l}\,\psi_{n}^{m}\left(x;l\right)\,. 
\label{eq:Polychrom}
\end{equation}
Orthogonality is inherited from Eq.~\ref{eq:OrthoNorm}, due to 
independence of spatial and spectral variables. 

The function basis, similar to that proposed in 
\citet{2010MNRAS.406.2433G}, 
is redefined here for minimization of the astrometric error 
from Sec.~\ref{sec:Astrometric-fit-error}, so that the fit expansion 
up to a given term provides a good estimate of the input signal 
position. 

Also, a \textit{common} polychromatic basis is used throughout 
our investigation. 
With respect to our previous work, this is more practical for 
application to the real data reduction, since the basis functions 
do not need to be generated anew for different spectral type stars. 
The model will thus be dependent on star colors only through the 
fit coefficient variation. 
The polychromatic basis is built for a near solar type source 
with blackbody spectrum corresponding to temperature $T = 6,000\, K$, 
which seems to be a reasonable practical trade-off from the results 
of our previous paper. 

\subsection{ Maximum likelihood fit approach } 
\label{sub:Maximum-likelihood-fit}
The fit approach is now cast in a framework similar to that of 
Sec.~\ref{sec:Astrometric-fit-error}. 

The detected signal $f\left(x_{k}\right)$ (the photo-center $\tau$
is not explicited in the next few steps) is approximated by the 
fit, represented by the expansion to the order $n$ 
using the basis functions ${\psi_{m}\left(x_{k}\right)}$: 
\begin{equation}
\tilde{f}_{n}\left(x_{k}\right) = \sum_{m=0}^{n}\, 
c_{m}\,\psi_{m}\left(x_{k}\right)\,. 
\label{eq:Expansion}
\end{equation}
The $\chi^{2}$ in Eq.~\ref{eq:ChiSquare}, taking advantage of 
Eq.~\ref{eq:FitError}, is then a function of the fit error 
to the same order: 
\begin{equation}
\chi_{n}^{2} = 
\sum_{k}\frac{h_{n}^{2}\left(x_{k}\right)}{\sigma_{k}^{2}}\,.
\label{eq:ChiSquare2}
\end{equation}
Taking the signal expansion from order $n$ to order $n+1$, the $\chi^{2}$ 
becomes 
\begin{equation}
\chi_{n+1}^{2} = 
\chi_{n}^{2}+c_{n+1}^{2}\sum_{k}\, 
\frac{\psi_{n+1}^{2}}{\sigma_{k}^{2}}-2c_{n+1}\sum_{k}\, 
\frac{f\,\psi_{n+1}}{\sigma_{k}^{2}}\,.
\label{eq:ChiSquare3}
\end{equation}
An ortho-normalization relationship can be imposed on the 
basis functions, introducing a modification to Eq.~\ref{eq:OrthoNorm} 
such that 
\begin{equation}
\left\langle \psi_{m}\,\psi_{n}\right\rangle = 
\sum_{k}\,\frac{\psi_{m}\left(x_{k}\right)\psi_{n}\left(x_{k}\right)} 
{\sigma_{k}^{2}} = \delta_{mn}\,. 
\label{eq:OrthoNorm2}
\end{equation}
This formulation of scalar product between functions over the 
domain implicitly gets rid of the sum in the mid term of 
Eq.~\ref{eq:ChiSquare3} (right-hand side). 

Then, we define the next expansion coefficient by requiring minimization
of the $\chi^{2}$: 
\begin{equation}
\frac{\partial\chi^{2}}{\partial c_{n+1}}=0\,,\label{eq:ReqNextCoeff}
\end{equation}
so that we get 
\begin{equation}
c_{n+1}=\sum_{k}\,\frac{f\left(x_{k}\right)\psi_{n+1}\left(x_{k}\right)}{\sigma_{k}^{2}}\,.\label{eq:ExpCoeffs}
\end{equation}

This expression corresponds to the projection of our input signal 
$f$ over the basis component $\psi_{n+1}$ using the 
definition of scalar product from Eq.~\ref{eq:OrthoNorm2}; 
it simplifies the above expression for $\chi^{2}$ to 
\begin{equation}
\chi_{n+1}^{2} = \chi_{n}^{2}-c_{n+1}^{2}\,, 
\label{eq:ChiSquare4}
\end{equation}
showing that each additional term effectively contributes to 
reduction of the fit discrepancy in the maximum likelihood sense. 
Also, the contribution of each term is limited, since the initial 
$\chi^{2}$ is also a limited quantity, thus ensuring the convergence 
of the fit process. 

However, the above formulation must be somewhat mitigated 
in order to be consistent with different signal instances, 
e.g. to be able to use a common basis to describe the 
different signal profiles, e.g. $f^{(r)}$ and $f^{(s)}$, 
from focal plane positions $r$ and $s$ associated to 
different instrumental response. 
The ortho-normalization relationship in Eq.~\ref{eq:OrthoNorm2} 
may not be expressed consistently for both signals, since 
in general $\sigma^{(r)}\neq \sigma^{(s)}$. 

Additionally, such formulation is apparently subject to 
degeneration in the photon limited (PL) case, where the variance 
equals the signal value in photon units: 
$\sigma_{k}^{2}(PL)=f\left(x_{k}\right)$. 
In particular, the denominator and numerator in the coefficient 
computation (Eq.~\ref{eq:ExpCoeffs}) or similar expressions 
would cancel out. 
The degeneration is solved in practice introducing an additional 
noise term $\sigma_{Noise}^{2}$, taking into account at least the 
readout noise and the shot noise related to the current background, 
i.e. 
\begin{equation}
\sigma_{k}^{2}=f\left(x_{k}\right)+\sigma_{Noise}^{2}\,. 
\label{eq:VarNoise}
\end{equation} 
However, this variance definition still depends on the individual 
exposure, i.e. on star magnitude, spectral type, background 
contribution, and current instrumental response. 

Our goal is to define an algorithm for signal fitting suitable 
to be applied in a straightforward way to a reasonably large set 
of input signals $\left\{ f^{(t)} \left(x_{k}\right) \right\}$, 
affected by different perturbations $(t)$.

\subsection{Choice of the weight function}
\label{sub:VarianceChoice}
The definition related to Eq.~\ref{eq:OrthoNorm2}, although optimal in 
the maximum likelihood sense with respect to a specific signal instance, 
associated to variance $\sigma_{k}^{2}$, is not suited to match 
\textit{at the same time} different cases within an ensemble of 
realistic signals, i.e. perturbations to the ideal system. 
The optimal basis for different signal instances are thus different, 
due to the dependence on the variance, 
which is impractical for implementation. 

In order to define a convenient common basis for a set of realistic
signal instances, a common weight distribution $w$ 
is adopted to replace the variance $\sigma$ from 
Eq.~\ref{eq:ChiSquare2} onward, mitigating 
the stringent requirement related to the maximum likelihood 
approach, and accepting some degradation with respect to the 
individual optimal fit. 
The goal is to achieve a good overall fit performance 
with a simple common model. 

In practice, we use an 
expression depending on a common reference function $F$ 
and an additive term $p$, corresponding to a pedestal applied 
to the signal, both chosen according to performance on the 
selected data set: 
\begin{equation}
w_{k} = \frac{F\left(x_{k}\right)+p}{1+p}\,. 
\label{eq:VarPedestal}
\end{equation} 
For convenience, $F$ is normalized to unity peak. 

It may be noted that the addition of a pedestal to the noiseless 
signal helps in balancing the relative contribution between the 
central, high slope region and the side pixels of the signal. 
Also, the normalization of the weight and reference function 
are of no consequence onto the astrometric error estimate (Eq.~\ref{eq:BiasEstim}), but they do affect the basis 
normalization (Eq.~\ref{eq:OrthoNorm2}). 

The choice of reference function $F$ and of pedestal value $p$ 
will be discussed in the framework of the simulations below, 
showing how they can be selected according to a convenient 
performance trade-off for a given data set. 

\section{ The Gaia study case }
\label{sec:GaiaSig}
Gaia \citep{2005ASPC..338....3P,2012Ap&SS.tmp...68D} 
is the ESA space mission devoted to high precision astrometry, 
to be launched in 2013. 
It will perform global astrometric measurements, with a parallax 
accuracy ranging from a few micro-arcsec (hereafter, $\mu as$) to 
few hundred $\mu as$ on individual stars, and a few milli-arcsec 
(hereafter, $mas$) for some asteroids, in the magnitude 
range between $V\simeq8\, mag$ and $V\simeq20\, mag$. 
Gaia operates in scanning mode, with continuous observation 
along a great circle moving over the full sky. 
It employs two telescopes with primary mirror size 
$L_\xi \times L_\eta = 1.45 \times 0.5\, m$, separated 
by a base angle of $106^{\circ}.5$, feeding a large common 
focal plane, composed of a mosaic of Charge Coupled Device (CCD) 
detectors operated in Time Delay Integration (TDI) mode. 

The Effective Focal Length value $EFL=35\, m$ corresponds to an 
optical scale $s=5".89/mm$: the $10\,\mu m$ detector pixel covers 
little less than $60\, mas$ on the sky. 
The Airy diameter of a diffraction limited optical image at a 
wavelength $\lambda=600\, nm$ is 
$2\lambda EFL / L_\xi \simeq 29\,\mu m \simeq 170\, mas$, 
i.e. of order of three pixels, or somewhat undersampled; 
the measured image is also affected by the source spectral 
distribution and realistic detection effects (e.g. MTF and 
operation). 
The end-of-mission accuracy goal, resulting from composition of 
the whole set of measurements, corresponds to a location error 
of the elementary exposure ranging from a few ten $\mu as$ for 
very bright stars to a few mas at the faint end, i.e. 
from $1/1,000$ to a few $1/100$ of the along scan size of the 
detector pixel. 

The readout mode depends upon the target brightness; binning is 
applied across scan (low resolution direction), to improve on 
signal to noise ratio (SNR) and reduce the data volume, for 
stars fainter than $V\simeq 13\, mag$. 
The output data is thus a {\em one-dimensional signal} in the high 
resolution direction, consistent with our problem formulation. 
\\ 
The signal is read out on a limited number of pixels (12 pixels 
for $13\lesssim V\lesssim 16\, mag$, six pixels for fainter stars), 
centered on the signal peak, providing the elementary 
exposure data. 
During the data reduction, the individual photo-centers are 
computed for the whole star sample, thus generating the inputs 
to the global astrometric solution \citep{Mignard2008}. 

The data reduction chain \citep{Busonero2010} 
includes the provisions for 
calibration based on self-consistency of the set of measurements, 
so that the instrumental parameters are estimated as well as 
astrometric and astrophysical parameters. 
The Gaia design benefits from the built-in differential measurement 
concept. 
Many stars are observed subsequently, with a time elapse of less 
than two hours, by both telescopes, over a limited region of the 
focal plane, i.e. nearby areas of the same CCDs. 
The field distortion contributions from either telescope are 
thus added to the measurements of each star, and may therefore 
be expected to average out in the repeated coverage of the whole 
sky. 
The pixel to pixel variation of CCD parameters is expected to 
be strongly reduced by TDI operation, thanks to averaging 
over each column. 
In practice, field distortion and other instrumental parameters 
are estimated from the measurements, and data are corrected 
to provide ``clean" individual photo-centers. 

It is possible to take advantage of the intrinsic different 
phase of many individual star observations with respect to 
the grid of the detector pixel array to retrieve the information 
on the underlying full resolution signal, i.e. a super-resolution 
PSF, as in \citet{2005A&A...436..373P}. 
Such synthetic high resolution signal can then be used for 
a convenient description of the ``true" PSF in the data reduction. 
For convenience, the fit may be defined on ``clean" data sets of 
sources not affected by exceeding peculiarity in their spatial 
structure, spectral distribution, temporal variability, astrometric 
behavior, and so on. 
The derived model can then be used, with proper cautions, for all 
objects observed, over the same period of time, through the same 
telescope, on that detector region. 
The model may also be retained until detection of a significant 
departure of the instrument response, based on monitoring its 
consistency with new data. 

Our simulations assume usage of such high resolution signals, 
representing the ``true" PSF, as input to the fit 
in different conditions. 

\begin{figure}
\includegraphics[width=75mm,height=50mm]{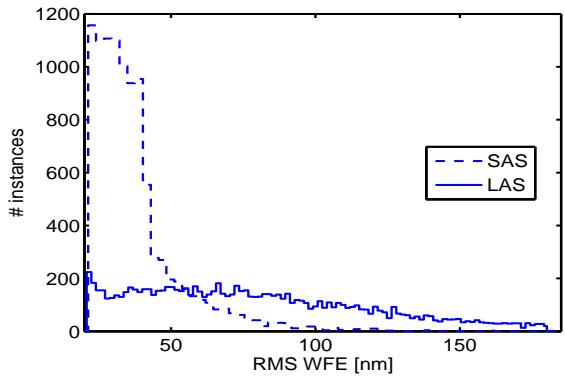}
\caption{\label{fig:RMS_WFE_LS}Distribution of RMS wavefront error for 
the two simulation data sets. }
\end{figure}

\begin{table}
\begin{center}
\caption{Mean and standard deviation of the RMS wavefront error over each 
simulation data set. \label{tab:RMS_WFE_LS}}
\begin{tabular}{ccc}
\tableline\tableline
RMS WFE & Mean {[}nm{]} & St. dev. {[}nm{]} \\ 
\tableline
SAS & 36.12 & 15.78 \\ 
LAS & 77.72 & 39.04 \\ 
\tableline
\end{tabular}
\end{center}
\end{table}

\section{Data sets and fit results}
\label{sec:Simulations1}
The goal of the simulations is to evaluate the practicality and 
performance of the above conceptual framework by application to 
different sets of realistic (i.e., not diffraction limited) signals. 

The first step of verification is focused on small perturbations, i.e. 
a set of signal instances affected by limited degradation with respect 
to the ideal case; the algorithm response can then be expected to be 
more easily understood. 
Larger perturbations are introduced in another data set to sample a 
more realistic range of variation. 
Hereafter, the two sets will be labeled as SAS and LAS, respectively 
for ``Small'' and ``Large'' Aberration Set; each contains $N=10,000$ 
independent signal instances 
$\left\{ f^{(n)}\left(x_{k}\right) = f(k;n) \right\}$ 
($n=1, \ldots, N$). 

We generate a set of perturbed images by construction of the optical 
point spread function (PSF), according to the diffraction integral 
description \citep{BornWolf}. 
The signal perturbations are introduced 
by means of independent wavefront errors, 
built by generation of random coefficients applied to Zernike 
polynomials of the pupil coordinates. 
The monochromatic PSF is composed with simple source spectra 
(blackbodies at given temperature) and detection effects 
(nominal pixel size, MTF and TDI operation; across scan binning) 
to build the effective detected signal, as in \citet{1998PASP..110..848G}. 
Such perturbations may represent at first order 
the effects of other disturbances, e.g. the charge transfer delay in a 
radiation damaged CCD, also modifying the signal profile. 
\begin{figure} 
\includegraphics[width=75mm,height=50mm]{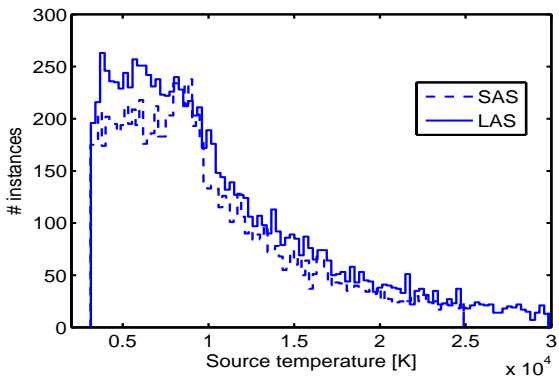}
\caption{\label{fig:Tsource_LS}Distribution of source temperature for the
two simulation data sets. }
\end{figure}

The spatial resolution of the simulated signals on the focal plane
is $1\,\mu m$ along scan and $3\,\mu m$ across scan, i.e. respectively
the high and low resolution coordinates of the instrument. The signal
is simulated and analyzed over a range of $\pm65\,\mu m$, corresponding
to $12\pm0.5$~pixels. The pupil sampling is $1\, cm$ in both directions.
The simulation is implemented in the Matlab environment. Some key
characteristics of the perturbation distribution associated to each
data set are summarized in Fig.~\ref{fig:RMS_WFE_LS}, which shows
the RMS WFE histogram for SAS (dashed line) and LAS (solid line),
and in Table~\ref{tab:RMS_WFE_LS}, reporting the RMS WFE mean and
standard deviation of both SAS and LAS. 
The instrument response quality of SAS data corresponds to a 
degradation of $\sim\lambda/20$ and rarely larger than $\lambda/10$, 
as expressed in terms of an effective wavelength $\sim700\, nm$, 
typical for intermediate type stars observed by Gaia. 
The LAS data are more evenly spread up to a degradation of 
$\sim\lambda/4$, roughly corresponding to Rayleigh's criterion 
for acceptable images.

Each signal instance is generated according to a different 
random blackbody temperature between 
$3,000\, K$ and respectively $25,000\, K$ (SAS) and $30,000\, K$ (LAS). 
The distributions, shown in Fig.~\ref{fig:Tsource_LS}, do not represent 
any special stellar population, but are just defined in order to span 
a reasonable range of spectral types. 
The spectral resolution is $20\, nm$.

In order to proceed to evaluation of the fit performance on either 
data set, it is necessary to choose the reference function $F$ 
defined in Sec.~\ref{sub:VarianceChoice}. 
The fit algorithm described in Sec.~\ref{sub:Maximum-likelihood-fit} is 
then applied to each signal instance. 
The fit quality is evaluated in terms of 
RMS signal discrepancy between input signal and fit, and of mean 
and RMS astrometric error. 
\\ 
{\em 
The astrometric error is evaluated for each signal instance 
according to Eq.~\ref{eq:BiasEstim} in the photon limit. 
}

The former aspect concerns the fit quality in the usual sense, 
whereas the latter two focus on the astrometric quality. 
The statistics over the data set is discussed as a function of 
the number of fitting terms and of the pedestal value $p$, ranging 
from $p=0.001$ (which corresponds to attributing negligible weight 
to the side pixels) to $p=1$ (signal central lobe weighted twice as 
much as the wings). 

\begin{figure}
\includegraphics[width=75mm,height=50mm]{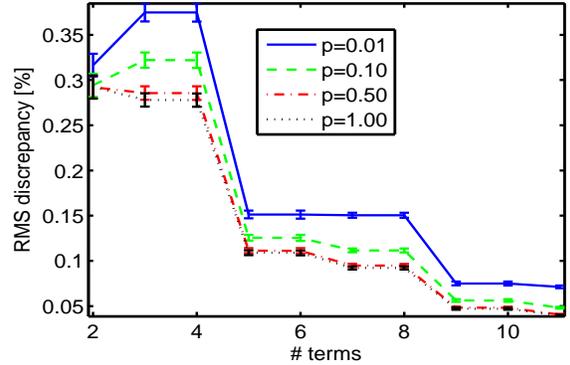}
\caption{\label{fig:PhotRMS_SAS1}Fit RMS discrepancy for SAS; average 
values (lines) and standard deviation (error bar) over the data set. }
\end{figure}

\begin{figure}
\includegraphics[width=75mm,height=50mm]{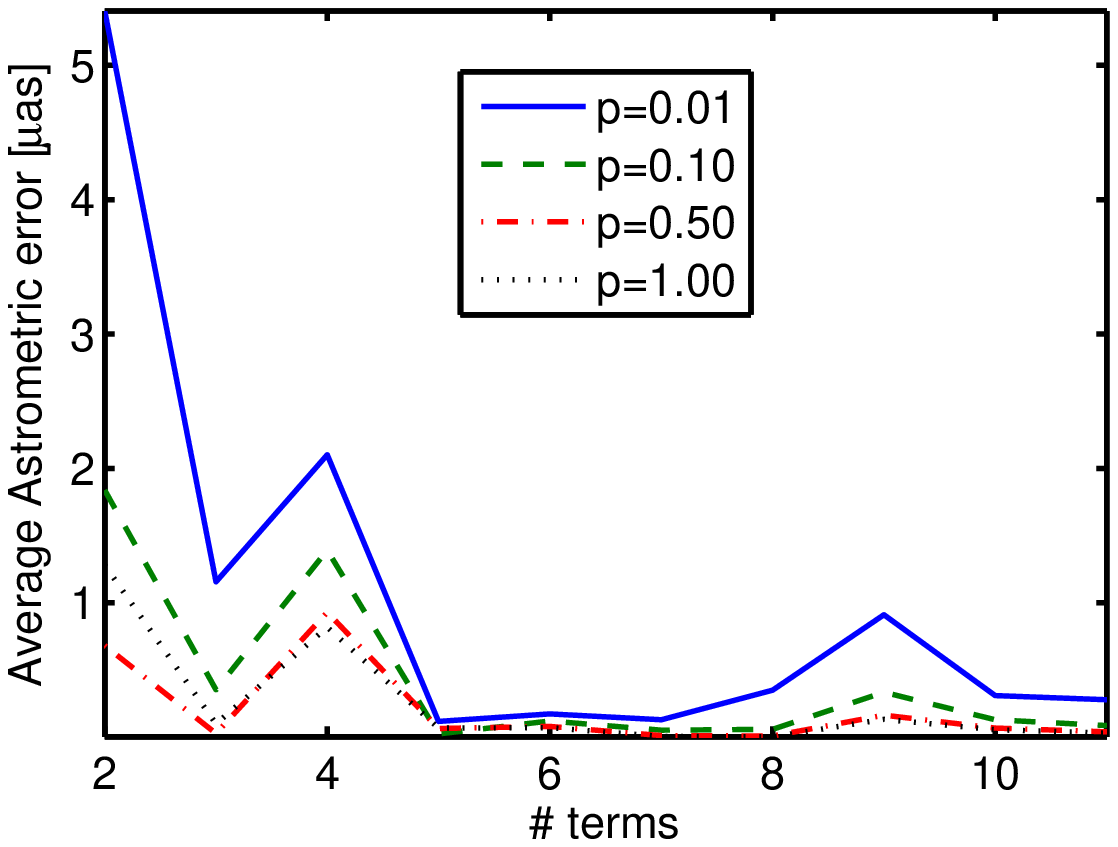}
\caption{\label{fig:AstrErrAve_SAS1}Average astrometric error over the 
SAS data, with pedestal values between 0.01 and one. }
\end{figure}

\begin{figure}
\includegraphics[width=75mm,height=50mm]{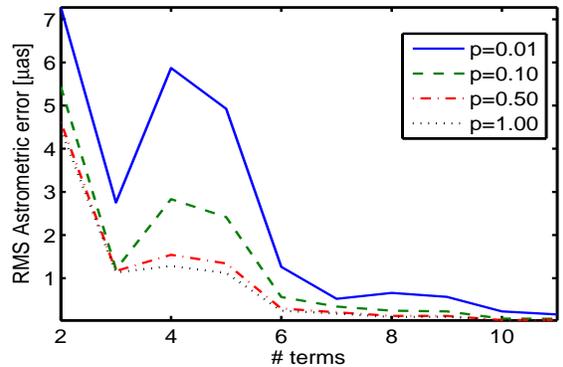}
\caption{\label{fig:AstrErrRMS_SAS1}RMS astrometric error over the SAS 
data, with pedestal values between 0.01 and one. }
\end{figure}

\subsection{Fit of SAS}
\label{sub:Fit-of-SAS}
The selected reference function $F$ for the variance is the 
polychromatic parent function from Eqs.~\ref{eq:Parent_function0} 
and \ref{eq:Polychrom}, geometrically scaled by $\sim6\%$ in order to 
match approximately the typical RMS width of the signal sample. 

The RMS fit discrepancy as a function of the number of fitting terms
is shown in Fig.~\ref{fig:PhotRMS_SAS1}, in terms of average (line) 
and standard deviation (error bar) over the data set for each pedestal 
value. 

The average and RMS astrometric errors as a function of the number
of fitting terms are shown respectively in Fig.~\ref{fig:AstrErrAve_SAS1}
and Fig.~\ref{fig:AstrErrRMS_SAS1}. 

The fit performance improves in general with increasing number of
fitting terms, but the pedestal has opposite effects on photometric
and astrometric errors. Increasing pedestal values are associated
to increasing photometric errors and decreasing astrometric errors. 

The RMS fit discrepancy is smaller for lower pedestal values, and/or 
for a larger number of fitting terms (which allow for better fitting 
the signal wings). 
Very low pedestal values appear to be detrimental to the 
astrometric performance, since both average and RMS astrometric 
errors exhibit significant fluctuations as a function of the number 
of fitting terms; larger pedestal values induce more monotonic 
behaviors. 
The sub-$\mu as$ variations for eight terms or more may be related to 
model limitations.

\subsection{Fit of LAS with SAS basis}
\label{sub:Fit-of-LAS}
The selected reference function $F$ for the LAS weight is based
on the average $\bar{f}$ of the signal sample 
$\left\{ f(k;n) \right\} $,
\begin{equation}
\overline{f}(x_k) = \frac{1}{N}\sum_{n}f(k;n)\,, 
\label{eq:LAS_Average}
\end{equation}
using in particular its symmetric component to avoid introducing bias
terms through the variance definition: 
\begin{equation}
F\left(x\right)=\frac{1}{2}\left[\overline{f}\left(x\right)+\overline{f}\left(-x\right)\right]\,.\label{eq:LAS_RefFun}
\end{equation}
The selected basis function is the same as for SAS, i.e. derived by
ortho-normalization of the non aberrated polychromatic signal and its
derivatives, according to Sec.~\ref{sec:Function-basis}. 

\begin{figure}
\includegraphics[width=75mm,height=50mm]{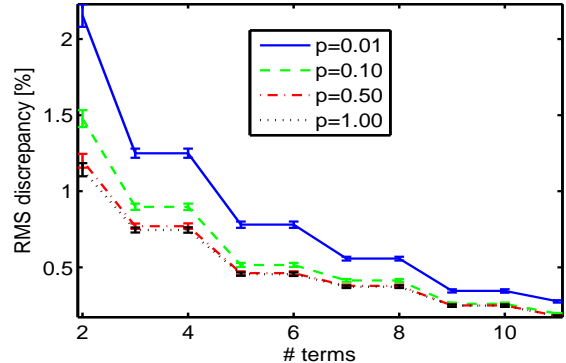}
\caption{\label{fig:PhotRMS_LAS1}Fit RMS discrepancy for LAS, with SAS basis;
average (lines) and standard deviation (error bar) over the data. }
\end{figure}

\begin{figure}
\includegraphics[width=75mm,height=50mm]{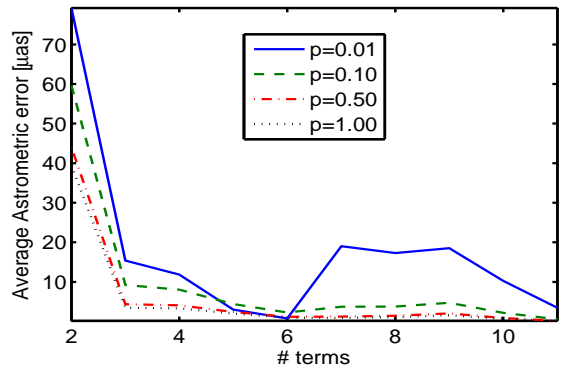}
\caption{\label{fig:AstrErrAve_LAS1}Average astrometric error over the LAS
data, using the SAS basis. }
\end{figure}

\begin{figure}
\includegraphics[width=75mm,height=50mm]{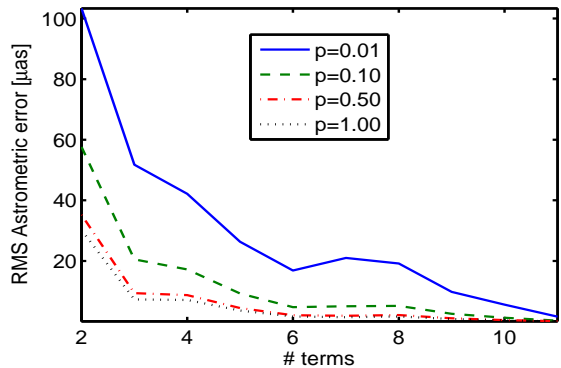}
\caption{\label{fig:AstrErrRMS_LAS1}RMS astrometric error over the LAS 
data, using the SAS basis. }
\end{figure}

The RMS fit discrepancy as a function of the number of fitting terms
is shown in Fig.~\ref{fig:PhotRMS_LAS1}, again in terms of average
(line) and standard deviation (error bar) for each pedestal value. 

The average and RMS astrometric errors as a function of the number 
of fitting terms are shown respectively in 
Fig.~\ref{fig:AstrErrAve_LAS1} and Fig.~\ref{fig:AstrErrRMS_LAS1}. 

As for the SAS case in Sec.~\ref{sub:Fit-of-SAS}, the fit performance 
has a tendency at improving with increasing number of fitting terms, 
with opposite effects of the pedestal on photometric and astrometric 
errors. 
Increasing pedestal values are associated to increasing photometric 
errors and decreasing astrometric errors. 
However, both photometric and astrometric errors are significantly 
larger than in the SAS case, because the larger perturbations require 
a larger number of fitting terms to account for the signal profile 
variations. Although it is possible to use the previous simple model 
for the current case of large signal perturbations, the need for many 
expansion terms to achieve $\mu as$ level fitting precision appears 
to be inconvenient for practical implementation. 

\begin{figure}
\includegraphics[width=75mm,height=50mm]{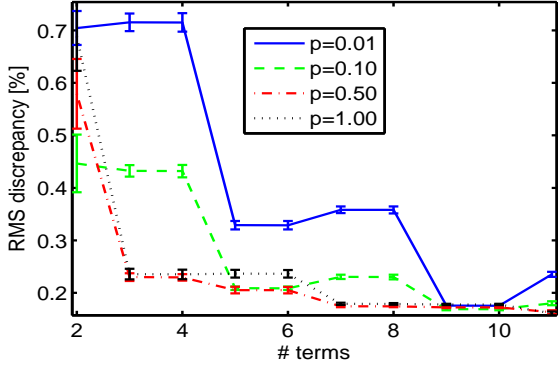}
\caption{\label{fig:PhotRMS_LAS2}Fit RMS discrepancy for LAS, with the 
modified basis; average (lines) and standard deviation (error bar) over 
the data. }
\end{figure}
\begin{figure}
\includegraphics[width=75mm,height=50mm]{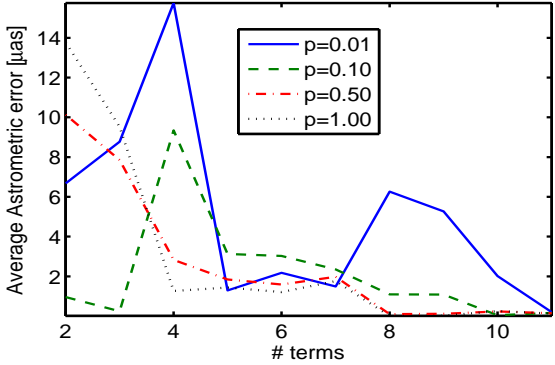}
\caption{\label{fig:AstrErrAve_LAS2}Average astrometric error over the 
LAS data, using the modified basis. } 
\end{figure}
\begin{figure}
\includegraphics[width=75mm,height=50mm]{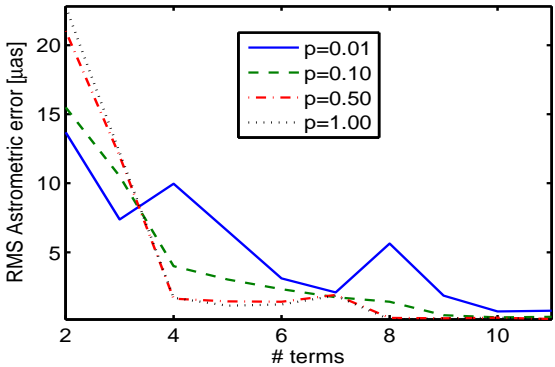}
\caption{\label{fig:AstrErrRMS_LAS2}RMS astrometric error over the 
LAS data, using the modified basis. } 
\end{figure}

\subsection{Basis tuning for LAS}
\label{sub:Basis-tuning}
The selected reference function is the same as above, but the basis
functions are modified to improve their similarity to typical LAS
data. In particular, the parent function becomes the symmetric average
signal used above as reference function $F$ for the variance, and the
subsequent basis terms are obtained by derivation and ortho-normality
according to the prescriptions in Sec.~\ref{sec:Function-basis}.
For convenience, the new parent function may be expanded in terms
of the SAS basis. 

The RMS fit discrepancy as a function of the number of fitting terms
is shown in Fig.~\ref{fig:PhotRMS_LAS2}, again in terms of average
(line) and standard deviation (error bar) for each pedestal value. 

The average and RMS astrometric errors as a function of the number
of fitting terms are shown respectively in Fig.~\ref{fig:AstrErrAve_LAS2}
and Fig.~\ref{fig:AstrErrRMS_LAS2}. 

The overall trend with number of fitting terms and pedestal value
is similar to the previous cases, but the improvement on fit quality
achieved by tuning the basis to the data set is significant. 
The photometric error of the fit drops to values comparable with the 
SAS case, i.e. RMS fit discrepancy below $0.5\%$ with few fitting terms 
and intermediate pedestal values. 
In the same conditions, the astrometric error drops to $<2\,\mu as$ 
values. 

Modification of the basis according to the current data set 
characteristics appears therefore to be convenient with respect 
to the achievable performance, and easily implemented for consistency 
with the current data. 

\begin{figure}[tr]
\includegraphics[width=75mm,height=50mm]{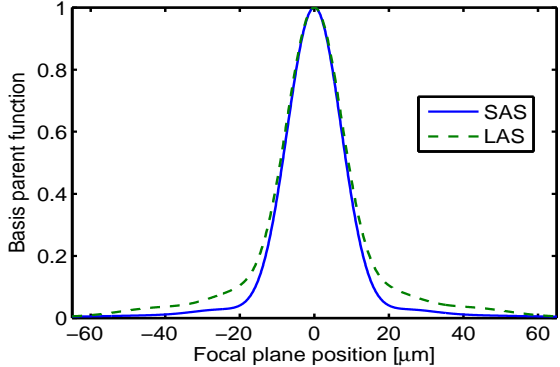}
\caption{\label{fig:ParentFunctions}Parent functions for SAS (solid 
line) and LAS (dashed line)}
\end{figure}

The two parent functions, i.e. the aberration free polychromatic 
signal (used for SAS) and the symmetric average signal (used for LAS 
throughout this section), are shown in Fig.~\ref{fig:ParentFunctions}, 
respectively with solid and dashed lines; it may be noted that the change 
in the function profile is small with respect to the peak value, 
and it may be described as a low intensity ``halo'' around the 
signal central peak, e.g. related to an average symmetric perturbation. 

\section{Spectral sensitivity}
\label{sec:Spectral-sensitivity}
The tolerance to the knowledge of the source effective temperature 
suggests that a limited number of different templates over the whole 
spectral range may be sufficient, thus alleviating the requirements 
of practical implementation within the data reduction system. 
The fit model performance as a function of the source temperature 
is a relevant issue, due to the potential risk of chromatic errors 
\citep{2007MNRAS.377.1337G,2006A&A...449..827B}. 
A preliminary indication on the model sensitivity can be deduced 
from the results of the previous simulation, e.g. splitting the 
astrometric error (for a given fitting choice) according to the 
instance temperature. 
\begin{figure}[tr]
\includegraphics[width=75mm,height=50mm]{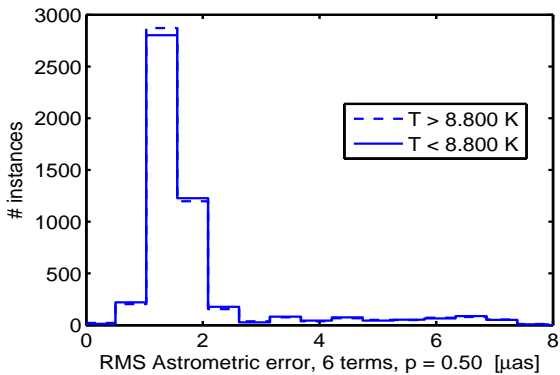}
\caption{\label{fig:AstrErr_T_LAS2}Histogram of the fit error 
distribution for LAS data, using the modified basis, above and below 
the median source temperature. }
\end{figure}
We select a temperature threshold $T_{T}=8,800\, K$, close to 
the median value of the LAS temperature distribution (solid line 
in Fig.~\ref{fig:Tsource_LS}), in order to have a comparable number 
of instances below and above such value. 

We select six terms for the fit, with a pedestal value 
$p=0.5$. 
The results are shown in Fig.~\ref{fig:AstrErr_T_LAS2}; the 
two histograms exhibit negligible difference. 
The fit errors seem thus to be mostly independent from the source 
temperature. 

In order to check the fit sensitivity to source temperature in 
more detail, similarly to the case of Sec.~2.2 of our previous 
paper, we simulate and analyze the data of a subset of 500 
perturbations, over the full range of source temperature. 
Using 20 temperature values, with uniform logarithmic spacing between 
3,000~K and 30,000~K, we get a set of 10,000 signal instances, as 
in Sec.~\ref{sec:Simulations1}, processed in a few hours by our 
hardware and software system. 

The signal perturbation is considered throughout this section as 
purely optical, and therefore completely described by diffraction; 
therefore, the variations with the source temperature may be different 
in practical cases for other kinds of disturbance. 
In case of perturbations with negligible dependence on the photon 
wavelength, the current results may thus represent a worst case. 

\begin{figure}
\includegraphics[width=75mm,height=50mm]{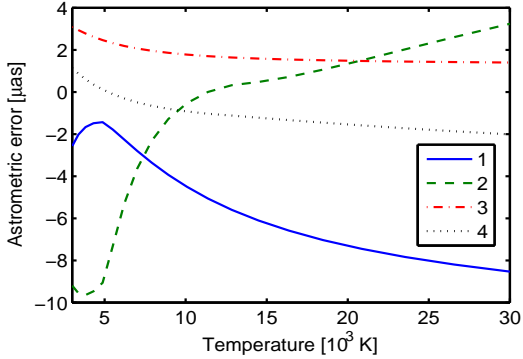}
\caption{\label{fig:SpectralBias}Astrometric error as a function of 
the source temperature for the first four LAS instances. }
\end{figure}

We retain six terms for expansion of the signal fit. 
The RMS fit discrepancy is practically independent from source 
temperature. The astrometric error as a function of the source 
temperature for the first four signal instances of LAS are shown 
in Fig.~\ref{fig:SpectralBias}. 
The slope variation is typically larger in the effective temperature 
range associated to near solar and later spectral types, whereas for 
higher temperatures the curves are smoother. 

\begin{figure}
\includegraphics[width=75mm,height=50mm]{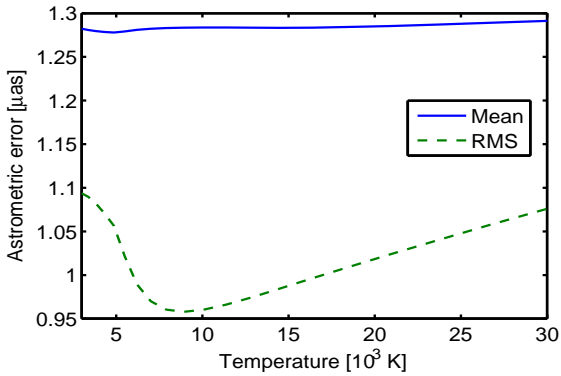}
\caption{\label{fig:AstrErr_col_LAS}Mean (solid line) and RMS (dashed line)
astrometric error vs. source temperature, averaged over the first
500 LAS instances. }
\end{figure}

The collective astrometric error is shown in 
Fig.~\ref{fig:AstrErr_col_LAS} vs. source temperature, 
respectively as mean (solid line) and RMS (dashed line) of the 
500 perturbed cases. 
The mean is basically constant, whereas the RMS features a shallow 
reduction for near solar type cases, possibly due to the temperature 
match between the source and the basis 
(Sec.~\ref{sec:Function-basis}). 
Both mean and RMS errors are close to the $1\,\mu as$ level, i.e. 
significantly smaller than the corresponding results of our previous 
paper. 
However, by comparison with Fig.~\ref{fig:SpectralBias}, the 
individual variation appears to be significant, suggesting that 
instrument response is relevant on the $\mu as$ range, requiring 
calibration and correction, with moderate requirements on 
knowledge of the source temperature. 

To verify the latter consideration, we simulate and analyze the 
data of the whole LAS data, upon application of a small variation 
to the source temperature, respectively $\pm1\%$ and $\pm5\%$, 
similarly to the case of Sec.~2.3 of our previous paper. 

\begin{figure}
\includegraphics[width=75mm,height=50mm]{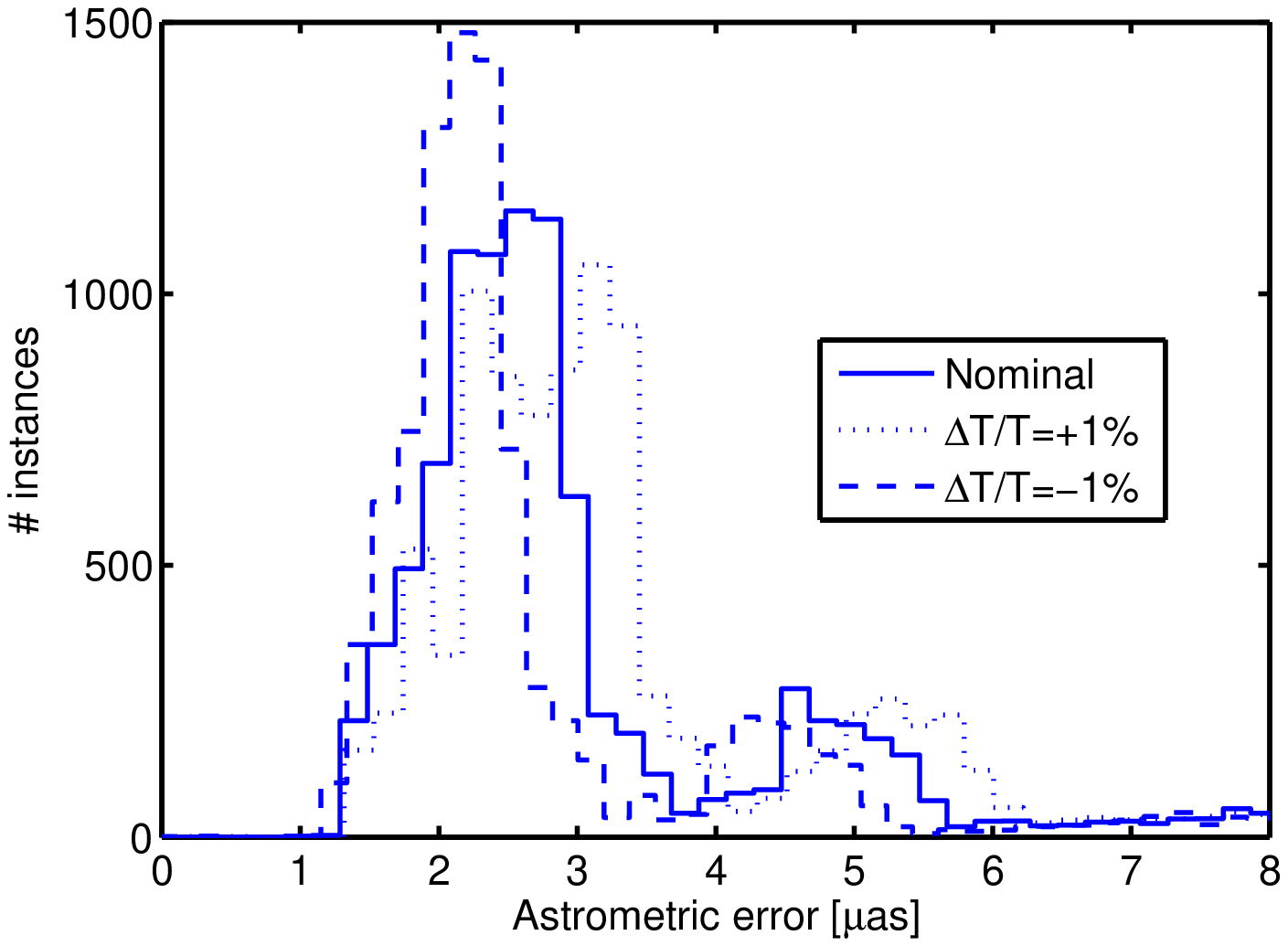}
\caption{\label{fig:AstrErr_dT1_LAS}Astrometric error on LAS, nominal 
and with $\pm1\%$ source temperature error. }
\end{figure}

\begin{figure}
\includegraphics[width=75mm,height=50mm]{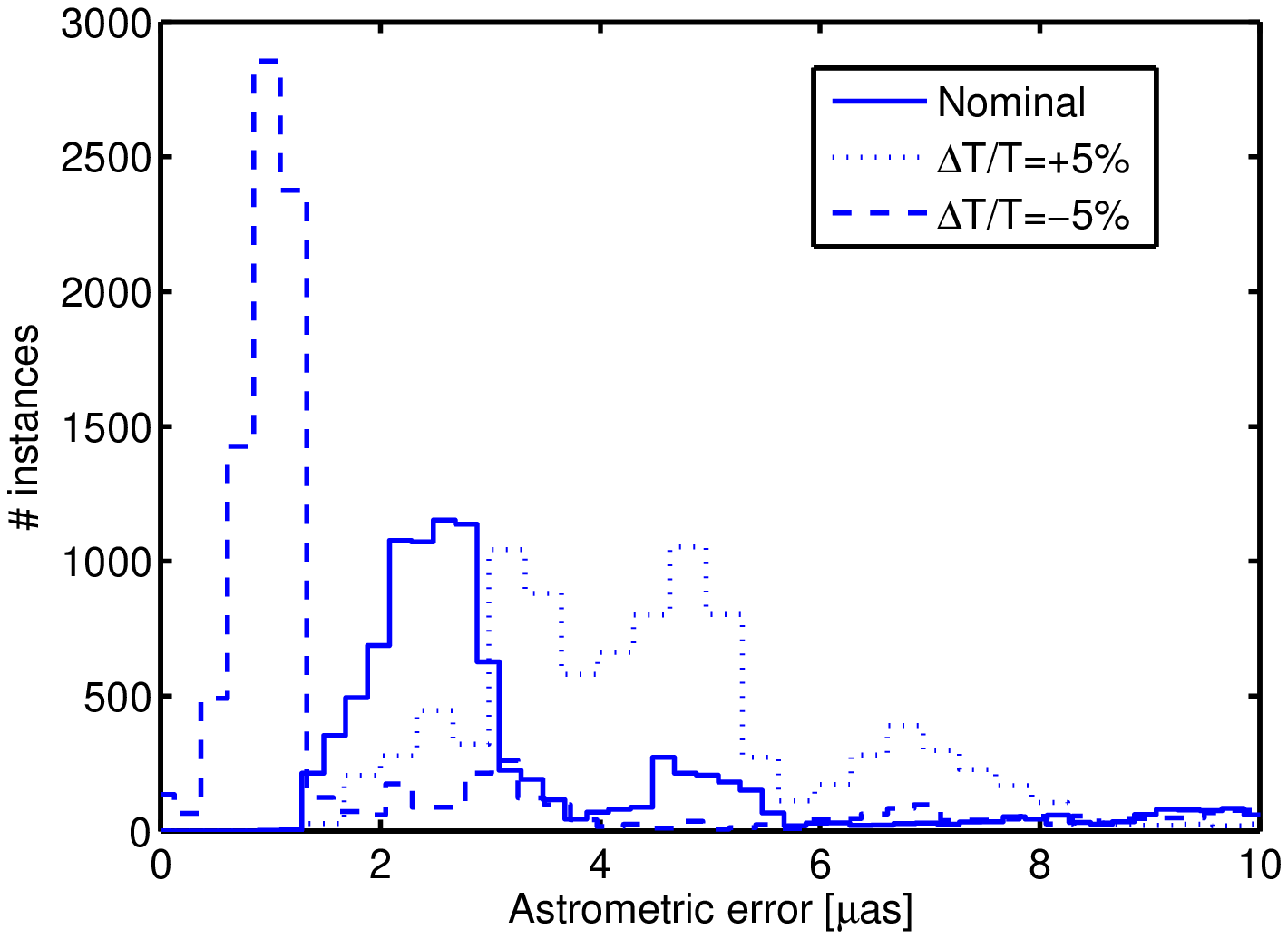}
\caption{\label{fig:AstrErr_dT5_LAS}Astrometric error on LAS, nominal 
and with $\pm5\%$ source temperature error. }
\end{figure}

This analysis may be relevant to two practical situations, 
related but not coincident. 
The former occurs during system calibration, in which the instrument 
response is being characterized, using the measurements on a set of 
different stars with similar, but not identical, detected spectrum. 
The latter concerns data reduction, in which the instrument response 
representation (encoded into the signal model represented by the 
fit) is used on all stars within the selected range of spectral 
types. 
The simulated model sensitivity can thus be used to select 
convenient spectral ranges for both computation and application 
of the signal model. 

The spectral distribution discrepancy between the reference signal 
(associated to the fit on nominal LAS data, limited to six terms) and 
the data generated with a given source temperature error induces an 
astrometric error, evaluated according to Eq.~\ref{eq:BiasEstim}. 
As above, the RMS fit discrepancy variation is marginal. 
In Fig.~\ref{fig:AstrErr_dT1_LAS} the distribution of astrometric 
error associated to the nominal LAS data and $\pm 1\%$ source 
temperature error is shown. 
Similarly, the astrometric error distributions related to $\pm 5\%$ 
source temperature error are shown in Fig.~\ref{fig:AstrErr_dT5_LAS}. 
The distributions exhibit a small systematic shift with respect to 
the nominal case, with the sign correlated to the temperature error. 
\begin{table}[tr]
\begin{center}
\caption{\label{tab:AstErr_dT}Mean and standard deviation of astrometric 
error associated to nominal LAS data and with $\pm1\%$ and $\pm5\%$ 
source temperature error }
\begin{tabular}{ccc}
\tableline\tableline
Astrometric error & Mean {[}$\mu as${]} & St. dev. {[}$\mu as${]}\\
\tableline
Nominal & 3.50  & 2.46 \\
$-1\%$ & 3.20  & 2.41 \\
$+1\%$ & 3.82  & 2.53 \\
$-5\%$ & 1.93  & 2.50 \\
$+5\%$ & 5.00  & 2.90 \\
\tableline
\end{tabular}
\end{center}
\end{table}

The mean and standard deviation of the astrometric error associated 
to the source temperature error is summarized in 
Table~\ref{tab:AstErr_dT}. 
The standard deviation over the sample ranges between $2.41\,\mu as$ 
and $2.90\,\mu as$, close to the dispersion of nominal data 
($2.46\,\mu as$); besides, the average value of astrometric error 
appears to be displaced by $\sim0.3\,\mu\, as$ per each $1\%$ variation 
in the source temperature.

\section{Discussion}
\label{sec:Discussion}
The estimate, from Eq.~\ref{eq:BiasEstim}, of the systematic 
astrometric error associated to the fit can be used to advantage 
in the development of a data reduction pipeline, and above all 
in providing a straightforward method to assess its behavior. 
The crucial point is that the bias introduced by the fit can be 
estimated in a straightforward way, either for introducing 
corrections in the data reduction, or to support in-depth 
analysis of the error distribution. 

It is not necessary to impose extremely high precision to the fit, 
e.g. using a large number of terms for expansion of the set of 
signals, since the error introduced to a given order can be easily 
assessed, and potentially used to correct the intermediate 
results. 
The trade-off can be set by the convenience of retaining limited 
errors over the whole focal plane, and for a suitably large 
range of source spectral variation, in order to keep the 
corrections small. 

The bias estimate can be applied to any fitting method, as it 
is derived from a general maximum likelihood approach. 
Its application to the basis proposed herein for 
signal expansion is a special case, in this sense. 
The performance is good, requiring order of six terms to reduce 
the RMS error to the $\mu as$ range over a large set of realistic 
signals. 

The signal profile is potentially different for each detector of 
a large focal plane, because of the variation on both optical image 
and device response. 
In the case of Gaia, it will be different for the two telescopes, 
at the $\mu as$ precision level. 
The signal fit must therefore be defined for at least 
$7\times 9 \times 2 = 126$ different profile instances, as many 
as the number of CCDs in the Gaia astrometric focal plane. 
A feeling of the implications on calibration and/or monitoring can be 
achieved in a straightforward way. 

The astrometric precision of an elementary exposure of an intermediate
magnitude and spectral type star is $\sim 300\,\mu as$. 
Therefore, in order to assess the instrument response at the 
$1\,\mu as$ level, i.e. to estimate the best fitting signal profile 
with such precision, we need to reduce the random error by a factor 
$\sim 300$, which requires averaging order of $300^{2}\simeq 10^{5}$ 
measurements.
Any astrometric CCD, with across scan size $\sim 5'$ and scanning 
rate $\sim 60"/s$, covers one square degree in about 12 minutes. 
Thus, with an average star density of $\sim 1,000$ stars per 
square degree, the required number of objects is collected in about 
20~hours. 

Actual application of the concept requires more detailed analysis, 
taking into account the real performance and stellar distribution. 
However, this back-of-the-envelope computation suggests that 
the science data may indeed be able to provide a self-calibration 
of the astrometric response of Gaia at the $\mu as$ level over 
a one day time scale. 
Similarly, the response variation across a CCD can be 
evaluated with comparable precision on a time scale of a few days. 

The spectral sensitivity is comparably small, since a $\pm 5\%$ 
error on the knowledge of the effective source temperature involves 
an astrometric error of order of $1\,\mu as$, negligible at the 
elementary exposure level for most stars, and partially averaged in 
the overall data set due to measurement over different focal plane 
positions of both telescopes. 
A set of $\sim 20$ templates, each computed for a different temperature 
mean value for a given focal plane position, can therefore provide an 
acceptable model for the spectral variation of the detected signal. 
This takes the number of fit instances required to describe the whole 
astrometric focal plane over the spectral range to $\sim2,500$. 
Using a six term expansion, this results in $\sim 15,000$ parameters. 

The instrument response variation over the field and with the source 
spectral type may represent a natural data set (with $\sim2,500$ 
instances) for definition of the best basis at a given time. 
However, the natural evolution of the instrument response during its 
life will introduce a corresponding variation of the optimal basis. 
At least two different strategies may be implemented. 
On one side, the signal basis may be updated from time to time, using 
within each period the best set of functions describing the current 
performance. 
On the other hand, the full set of measurements can be considered as a 
unique statistical sample to define the overall best common basis. 
The latter approach has the benefit of a common signal representation 
for the whole set of data, potentially at the expense of some increase 
of the astrometric noise. 
The impact on overall systematic errors (e.g. regional astrometric 
errors for the Gaia catalog) may be evaluated by further large 
scale simulations.

\section{Conclusions}
\label{sec:Conclusions}
We propose an estimate of the fit quality of one-dimensional signals 
for astrometry, based on a maximum likelihood approach. 
Building on this framework, and on a simple definition of the 
function basis, we derive a fitting model easily tailored to a 
given set of signals. 
Its performance is evaluated by simulation over a set of realistic 
signal instances, affected by significant perturbation levels. 
The fit quality is analyzed as a function of the number of terms 
used for signal expansion, taking into account not only the 
RMS discrepancy with respect to the input signal, but above all 
the astrometric error associated to the fit. 

The maximum likelihood fit provides micro-arcsec astrometric errors, 
and RMS fit discrepancy of a few $0.1\%$, using six terms. 
The basis functions can be conveniently tuned to a selected set of 
signal profiles, e.g. to match the actual instrument response and/or 
its evolution. 
The sensitivity to a priori knowledge of the source spectral distribution 
is low, allowing usage of a limited number of templates. 

Some implications on monitoring and calibration of an astrometric 
instrument using the proposed signal expansion method are discussed, 
with reference to the typical parameters of Gaia. 

\acknowledgments
The activity is partially supported by the contract ASI I/058/
10/0. The authors' understanding of the subject and its implications
benefits from discussions with M. Shao and with M.
Lattanzi. The paper clarity significantly improved thanks to
the amendments suggested by the referee.

\clearpage

\end{document}